# Electromagnetic Isolation Induced by Time-Varying Metasurfaces: Non-Reciprocal Bragg Grating

Davide Ramaccia, *Senior Member, IEEE*, Dimitrios L. Sounas, *Senior Member, IEEE,*
Angelica V. Marini*, Student Member, IEEE,* Alessandro Toscano, *Senior Member, IEEE,*
and Filiberto Bilotti, *Fellow, IEEE*

*Abstract*—In this letter, we propose a magnet-less non-reciprocal isolating system based on time-varying metasurfaces. Two parallel time-varying metasurfaces, one for frequency up-conversion and one for down-conversion by the same amount, are used for realizing a region of space where incident waves from opposite directions experience an opposite Doppler frequency shift. As a result, any device within this region becomes sensitive to the illumination direction, exhibiting a different scattering response from opposite directions and thus breaking reciprocity. Very importantly, thanks to the opposite frequency shift of the metasurfaces, the frequency of the transmitted electromagnetic field is the same as for the incident one. Here, we demonstrate this general approach by using a Bragg grating as the device between the time-varying metasurfaces. The combined structure of the metasurfaces and the grating exhibits different transmission and reflection properties for opposite illumination direction, thereby realizing an isolator. More broadly, this letter presents a strategy for converting any conventional electromagnetic device to a non-reciprocal one by placing it between two time-varying metasurfaces. This approach opens the door to several new non-reciprocal components based on thin and lightweight metasurfaces, which are simpler to realize compared to their volumetric counterparts.

## I. Introduction

ELECTROMAGENTIC isolation is achieved when the propagation of an electromagnetic (EM) wave is allowed in one direction but not in the opposite one [1]. This is a non-reciprocal response and it is typically achieved through non-reciprocal materials, such as magnetized ferrite, which have asymmetric permittivity or permeability tensors, thereby breaking the conditions of the Lorentz reciprocity theorem [2]. However, in the last few years, it has been demonstrated that non-reciprocity can also be achieved by the so-called spatiotemporally modulated (STM) metamaterials, artificial electromagnetic materials whose refractive index changes periodically in space and time, effectively rendering them moving with a certain velocity in one specific direction, thereby allowing breaking reciprocity [3], [4]. Some non-reciprocal magnet-less devices that have been proposed in this context include isolators [3], [5], circulators [6], [7], non-reciprocal antennas [8]–[11] and Doppler cloaks [12], [13]. However, the practical realization of space-time modulated metamaterials is not a straightforward task [4], [14].

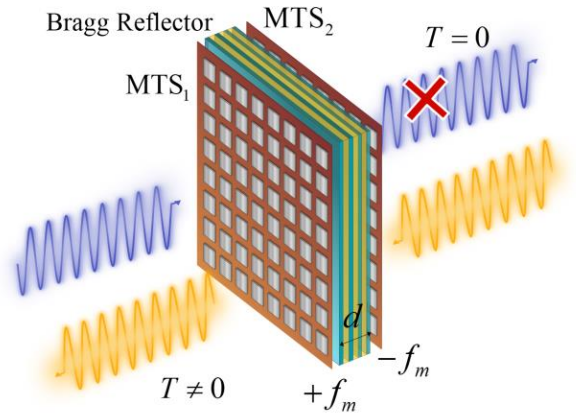

Fig. 1. Graphical representation of the functionality of the proposed magnet-less isolating system based on time-varying metasurfaces (TV-MTS). The system consists of two TV-MTSs and a reciprocal filtering selective structure between them. The overall non-reciprocal behavior is achieved by the opposite frequency conversion of the illuminating electromagnetic wave induced by the two metasurfaces. From the source and receiver point-of-view, the system exhibits a non-reciprocal response at the operative frequency $f_0$.

More recently, time-varying metasurfaces (TV-MTSs) have been investigated as a platform for achieving the same control on the scattered fields as bulk STM metamaterials but in a simpler and electrically thin electromagnetic structure, for both reflection [15], [16] and transmission [17], [18]. These works

Manuscript received March 31, 2020.
Davide Ramaccia, Angelica Marini, Alessandro Toscano and Filiberto Bilotti are with the Department of Engineering at ROMA TRE University, Via Vito Volterra 62, 00146 Rome, Italy (e-mail: davide.ramaccia@uniroma3.it).
Dimitrios L. Sounas is with the Department of Electrical and Computer Engineering, Wayne State University, Detroit, MI 48202, USA (e-mail: dsounas@wayne.edu).

The work has been developed in the framework of the activities of the research contract MANTLES, funded by the Italian Ministry of Education, University and Research as a PRIN 2017 project (protocol number 2017BHFZKH).

have shown that by linearly increasing or decreasing in time the phase of the transmission or reflection coefficient through a proper temporal modulation of the effective surface electric and magnetic impedances, a frequency shift can be realized to impinging waves. Although a time-varying metasurface may exhibit non-reciprocal response when also spatial modulation is applied [19], [20], in the case of a purely time-modulated metasurface, the transmitting response is independent on the illumination direction, *i.e.*, it induces the same frequency (up- or down-) conversion regardless the normal illumination direction, allowing us to classify it as a reciprocal component [17]. As a result, a time-varying metasurface cannot be used in a stand-alone configuration for achieving non-reciprocity.

In this letter, we propose a non-reciprocal isolator by combining the frequency shifting properties of time-varying metasurfaces with the frequency dispersive properties of *unmodulated,* and therefore reciprocal, structures. The aim of this paper is twofold: 1) *introduce a novel approach for designing non-reciprocal devices,* leveraging the frequency conversion properties of TV-MTSs, and 2) *demonstrate the proposed approach in the simple scenario of an isolator realized through Bragg grating between two ideal time-varying metasurfaces.* The proposed approach overcomes the limitations exhibited by previous ones based on bulk STM metamaterials, which are challenging to implement and have efficiency limitations [3], [5]. The non-reciprocity based on TV-MTSs may open the door to a new family of non-reciprocal devices based on lightweight, electrically thin, and easily controllable artificial surfaces.

The paper is organized as follows: in Section II, we present the geometry and operative principle of the proposed isolating system, Section III is devoted to the design of the system for operating at 10 GHz and discussion of the numerical results. Finally, in Section IV some conclusions are drawn, reporting the importance of the work or suggesting other potential applications and extensions.

## II. Electromagnetic Isolator based on Time-Varying Metasurfaces

### A. Geometry

In Fig. 2 we report the geometry of the electromagnetic isolator considered in this work. The system is illuminated by a normally impinging monochromatic *y*-polarized plane wave at frequency $f_0$ propagating along the *z*-direction. It consists of a conventional Bragg grating between two time-varying metasurfaces, represented by the blue and red dashed lines in Fig. 2, which are infinite along the *x*- and *y*-directions. The Bragg grating consists of dielectric slabs with permittivity $\varepsilon_{r1}$, $\varepsilon_{r2}$ and thickness $d_1$, $d_2$, respectively, and like the metasurface is also infinite along the *x*- and *y*-directions. The permittivity contrast between the two dielectric slabs and the number of layers that make up the Bragg grating defines its transmission and reflection responses as a function of frequency, according to the theoretical analysis reported in several textbooks (e.g., [21]). Two time-varying metasurfaces are placed before and after the Bragg grating, and they are modeled here as two electromagnetic transition conditions that impart a linearly increasing (for up-conversion) or decreasing (for down-conversion) phase versus time to the transmission coefficient, in order to induce the desired frequency shift to the transmitted fields. This simplified scenario allows us to focus our attention to the operation principle of the proposed approach for achieving isolation with time-varying metasurfaces.

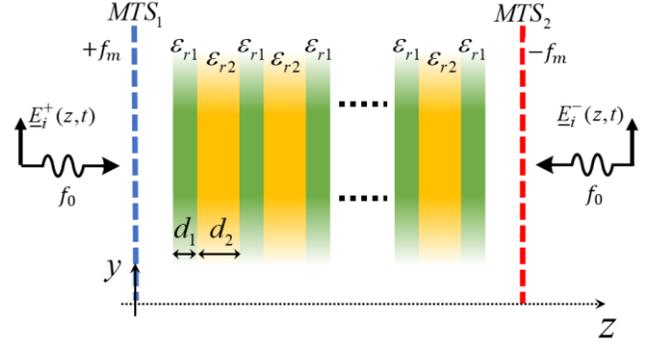

Fig. 2. Geometrical sketch of the proposed non-reciprocal Bragg-grating. The Bragg grating consists of a periodic alignment of a pair of infinitely extended dielectric slabs with permittivity $\varepsilon_{r1}$, $\varepsilon_{r2}$ and thicknesses $d_1$, $d_2$, respectively. The two metasurfaces located at the beginning and end of the Bragg grating are modeled as a uniform time-varying surface electric and magnetic impedances. The system is illuminated by a TEM plane wave at the frequency $f_0$.

### B. Operation principle

To illustrate the operation principle of the proposed system, we first report the typical transmissivity response of a Bragg grating around its bandgap edge, where frequency dispersion is maximum and transmission changes from low to high values (Fig. 3).

In Fig. 3a, the isolating system is illuminated from the left. If the illumination frequency $f_0$ is close to the peak transmission frequency $f_+$ of the Bragg grating, a small modulation frequency can be used to upconvert the illuminating field from $f_0$ to the frequency $f_+ = f_0 + f_m$, resulting in high transmission through the grating (Fig. 3a-left). The transmitted field, still at the frequency $f_+$, interacts now with the down-converting metasurface, restoring the frequency to its original value $f_0$ (Fig. 3a-right).

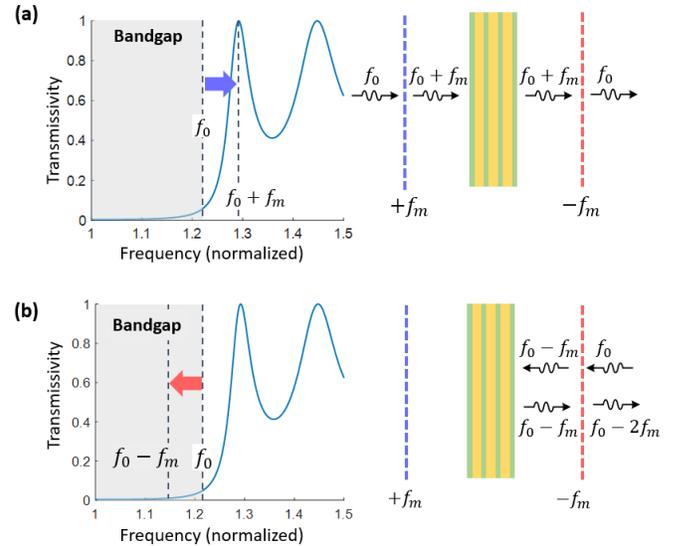

Fig. 3. Operation principle of the proposed isolating system based on Bragg grating and two time-varying metasurfaces: (a) the isolating system is illuminated from the left, where the up-converting TV-MTS shifts the illumination frequency within the transmission band of the Bragg grating; (b) the isolating system is illuminated from the right, where the down-converting TV-MTS shifts the illumination frequency deeper in the bandgap.

Consider now the case of illumination from the right (Fig. 3b). In this case, the electromagnetic field impinges on the red-shifting metasurface and therefore its frequency is converted to $f_- = f_0 - f_m$, which is now within the bandgap of the Bragg grating, resulting in high reflection. The reflected field interacts again with the time-varying metasurface, experiencing a second down-conversion to the frequency $f_0 - 2f_m$. In this case, the transmitted field is zero, whereas the reflected field is at a different frequency with respect to the source frequency, which is necessary for the system to obey passivity at $f_0$.

It is clear that in general the isolation level of the non-reciprocal system described above strongly depends on the frequency dispersion of the frequency selective structure: the sharper the transition from the forbidden to the transmitting band is, the higher the isolation is for a given modulation frequency $f_m$, or alternatively speaking, the lower the required $f_m$ is to achieve a desired isolation. The latter is especially important for easily satisfying the realization constrains in terms of modulation frequency [15], [20], [22], [23].

## III. DESIGN AND NUMERICAL VERIFICATION

In the following, we explain the design of the proposed system for operation in the X-band at frequency $f_0 = 10 GHz$. The Bragg grating has been designed by using two dielectric slabs with permittivity $\varepsilon_{r1} = 10$ and $\varepsilon_{r2} = 2$, respectively, and thicknesses $d_1 = 3.0mm$ and $d_2 = 6.7mm$, respectively. The number of layers is 21 (10 for each permittivity value, plus one with higher permittivity at the beginning to restore the symmetry of the device [21]). In Fig. 4, we report the scattering coefficients $S_{xy}$ with $x,y=1,2$ of the designed Bragg grating as a function of frequency. The scattering parameters have been computed though numerical simulations in CST studio suite [24]. The frequency range [9.8 GHz; 10.2 GHz] was selected so that within it the Bragg grating exhibits the transition from the forbidden frequency band ($f < 10GHz$), where the transmission is $S_{21} < -10dB$, to the transmitting frequency band ($f > 10GHz$), where the transmission is $S_{21} \geq -10dB$.

The transmission coefficients of the complete system, i.e. Bragg grating and the two time-varying metasurfaces, have been computed though a set of full-wave numerical simulations based on a 1D Finite Difference Time Domain (1D-FDTD) technique. In the simulation, the frequency shift induced by the time-varying metasurfaces has been successfully achieved by modeling them as two electrically thin matched modulated films of thickness $d$ with equilibrated periodic temporal modulation [25]. In this case, the refractive index $n^\pm(t)$ and impedance $\eta^\pm(t)$ of each film are:

$$n^\pm(t) = n_0 \sqrt{\mu_r^\pm(t)\varepsilon_r^\pm(t)} \qquad (1)$$

$$\eta^\pm(t) = \eta_0 \sqrt{\mu_r^\pm(t)/\varepsilon_r^\pm(t)} \qquad (2)$$

where the superscript "$\pm$" identifies the up-/down- converting thin film, $n_0, \eta_0$ are the free-space refractive index and impedance, respectively, and $\varepsilon_r^\pm, \mu_r^\pm$ are the time-varying relative permittivity and permeability of the thin film. Equilibrated periodic temporal modulation requires $\varepsilon_r^\pm(t) = \mu_r^\pm(t)$, that, according to (2), returns an always-matched time-varying film $\eta^\pm(t) = \eta_0$. As for the refractive index $n^\pm(t)$, it must linearly increase (or decrease) versus time for imposing a linear variation of the phase of the transmission coefficient. This can be achieved by selecting:

$$\varepsilon_r^\pm(t) = \mu_r^\pm(t) = \xi_r [1 \pm \delta f_m t], \qquad (3)$$

where $\xi_r = \varepsilon_r, \mu_r$ (with $\varepsilon_r = \mu_r$), $f_m$ is the desired frequency shift, $t$ is time, and $\delta$ is the modulating coefficient properly selected for achieving a periodic $2\pi$ phase delay (or phase advance) across the thin film of the transmission coefficient after each modulation period $T_m = 1/f_m$. It is easy to show that under these conditions the frequency shift is equal to $(1/2\pi)d\varphi^\pm/dt = \pm f_m$, where $\varphi^\pm = k_0 n^\pm L$ is the transmission phase, $k_0$ is the free-space wavevector and $L$ the length of the slab. It is worth noting that, considering phase wrapping, i.e., phases separated by $2\pi$ lead to the same result, the time-modulated phase linearly increases within a time interval $T_m = 1/f_m$ and then returns to its initial values, following a sawtooth temporal profile. The same profile can also be used for the time-varying permittivity/permeability, instead of the strictly linear one in (1). It is important to highlight that the proposed implementation of the time-varying transmitting metasurfaces is used only for its relatively easy implementation in a numerical simulation. For a more realistic implementation, one can follow the approaches in [17].

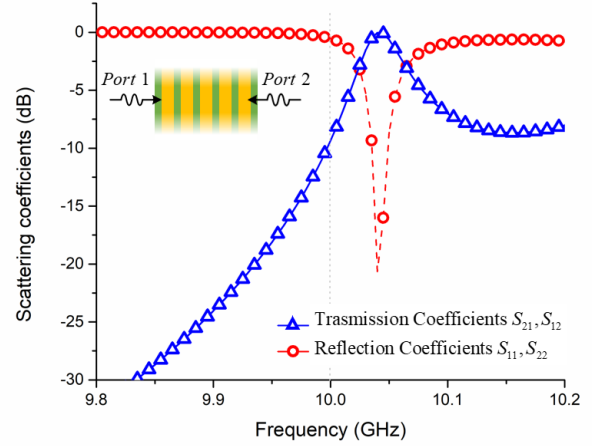

Fig. 4. Numerically computed scattering parameters of the designed Bragg grating without the presence of the time-varying metasurfaces. At the operative frequency $f_0 = 10GHz$, the transmission is -10dB.

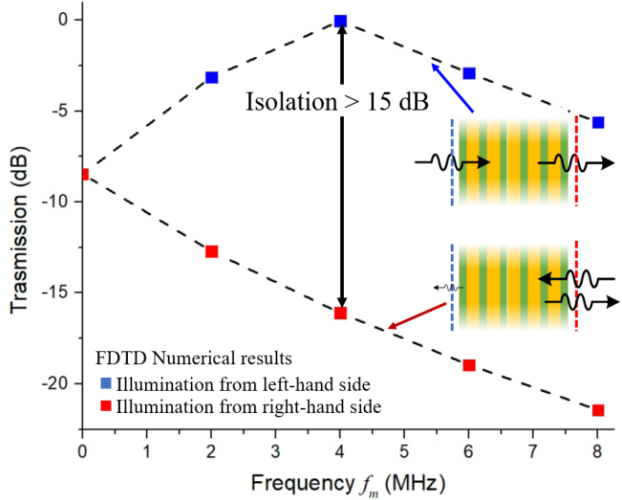

Fig. 5. Numerically computed amplitude of the transmission coefficient at the operative frequency $f_0 = 10 GHz$ as a function of the modulation frequency $f_m$ when the system is illuminated from left (blues squares) and from right (red squares).

The non-reciprocal Bragg grating has been simulated for five different modulation frequencies, i.e. $f_m = [0, 2, 4, 6, 8] MHz$ and for both illumination directions. The excitation signal is a narrow band gaussian pulse with center frequency $f_0 = 10 GHz$. The amplitude of the transmitted fields at $f_0$ have been evaluated via Fourier transform of the time-domain transmitted signal after the output metasurface. In Fig. 5 we report the transmission level in dB for different modulation frequencies and for the two opposite illumination directions. When the system is illuminated from the left-hand side, the up-converting metasurface shifts the excitation signal towards the frequency at which the transmission is maximum, as shown in Fig. 4, blue solid curve. The maximum is reached when the modulating frequency $f_m = 4 MHz$, and further increasing the modulation frequency leads to a reduction of transmission, due to the narrow width of the grating's transmission band.

On the contrary, when the system is illuminated from right, the amplitude of the transmitted field at $f_0$ decreases for higher values of the modulation frequencies, since the illumination signal is shifted deeper and deeper inside the bandgap of the Bragg grating. The maximum isolation exhibited with the system is about 17dB, as shown in Fig. 5, fully validating the operation of the electromagnetic isolator based on time-varying metasurfaces driven by small modulation frequencies. Higher values can be achieved if a grating or another filtering structure with a higher Q-factor is employed. The Q-factor of the filtering element realizing the metasurface-based isolator is also directly responsible for the operation bandwidth of the system, since it imposes a limit on the frequency range that can pass through the proposed non-reciprocal system. This limitation can be significantly relaxed if the response of the filtering element exhibits a much wider pass-band behaviour and a steep roll-off from the stop- to the passband. Another possibility to increase the isolation is to move the operation frequency $f_0$ farther from the peak transmission frequency $f_+$. However, this case requires higher modulation frequencies, imposing an upper bound to how far $f_0$ is from $f_+$ related to realization constrains [15].

## IV. CONCLUSION

In this letter, we have proposed a magnet-less non-reciprocal isolating system based on time-varying metasurfaces. The frequency conversion capability of time-varying metasurfaces has been exploited for realizing a Doppler effective region, where the propagating electromagnetic fields are at a different frequency with respect to the source one depending on the propagation direction. The non-reciprocal response has been obtained by inserting a conventional Bragg grating between the two time-varying metasurfaces. Despite the source frequency is the same for both propagation directions, the Bragg grating is excited by two different frequencies when illuminated from the two opposite sites, as if it were in apparent motion with respect to the source. In general, we have demonstrated that it is possible to overcome the limitations shown by magnet-less isolators based on STM metamaterials, by just introducing a conventional reciprocal electromagnetic between two time-varying metasurfaces, thus converting it to a non-reciprocal device. The non-reciprocity based on TV-MTSs may open the door to a new family of non-reciprocal devices based on light weight, electrically thin, and easily controllable artificial surfaces.

## REFERENCES

[1] D. M. Pozar, *Microwave engineering*. Wiley, 2012.
[2] A. Gurevich and G. Melkov, *Magnetic Oscillations and Waves*. CRC Press, 1996.
[3] Z. Yu and S. Fan, "Complete optical isolation created by indirect interband photonic transitions," *Nat. Photonics*, vol. 3, no. 2, pp. 91–94, Jan. 2009.
[4] S. Qin, Q. Xu, and Y. E. Wang, "Nonreciprocal Components With Distributedly Modulated Capacitors," *IEEE Trans. Microw. Theory Tech.*, vol. 62, no. 10, pp. 2260–2272, Oct. 2014.
[5] N. Chamanara, S. Taravati, Z.-L. Deck-Léger, and C. Caloz, "Optical isolation based on space-time engineered asymmetric photonic band gaps," *Phys. Rev. B*, vol. 96, no. 15, p. 155409, Oct. 2017.
[6] A. Kord, D. L. Sounas, and A. Alu, "Magnet-Less Circulators Based on Spatiotemporal Modulation of Bandstop Filters in a Delta Topology," *IEEE Trans. Microw. Theory Tech.*, vol. 66, no. 2, pp. 911–926, Feb.

[7] N. A. Estep, D. L. Sounas, J. Soric, and A. Alù, "Magnetic-free non-reciprocity and isolation based on parametrically modulated coupled-resonator loops," *Nat. Phys.*, vol. 10, no. 12, pp. 923–927, Dec. 2014.

[8] D. Ramaccia, D. L. Sounas, A. Alù, F. Bilotti, and A. Toscano, "Nonreciprocal Horn Antennas Using Angular Momentum-Biased Metamaterial Inclusions," *IEEE Trans. Antennas Propag.*, vol. 63, no. 12, 2015.

[9] D. Ramaccia, D. L. D. L. Sounas, A. Alu, F. Bilotti, and A. Toscano, "Nonreciprocity in antenna radiation induced by space-time varying metamaterial cloaks," *IEEE Antennas Wirel. Propag. Lett.*, vol. 17, no. 11, pp. 1968–1972, Nov. 2018.

[10] S. Taravati and C. Caloz, "Space-time modulated nonreciprocal mixing, amplifying and scanning leaky-wave antenna system," in *2015 IEEE International Symposium on Antennas and Propagation & USNC/URSI National Radio Science Meeting*, 2015, pp. 639–640.

[11] Y. Hadad, J. C. Soric, and A. Alu, "Breaking temporal symmetries for emission and absorption," *Proc. Natl. Acad. Sci.*, vol. 113, no. 13, pp. 3471–3475, Mar. 2016.

[12] D. Ramaccia, D. L. Sounas, A. Alù, A. Toscano, and F. Bilotti, "Doppler cloak restores invisibility to objects in relativistic motion," *Phys. Rev. B*, vol. 95, no. 7, p. 075113, Feb. 2017.

[13] D. Ramaccia, D. Sounas, A. Alu, A. Toscano, and F. Bilotti, "Advancements in Doppler cloak technology: Manipulation of Doppler Effect and invisibility for moving objects," in *2016 10th International Congress on Advanced Electromagnetic Materials in Microwaves and Optics (METAMATERIALS)*, 2016, pp. 295–297.

[14] S. Taravati, "Giant Linear Nonreciprocity, Zero Reflection, and Zero Band Gap in Equilibrated Space-Time-Varying Media," *Phys. Rev. Appl.*, vol. 9, no. 6, p. 064012, Jun. 2018.

[15] D. Ramaccia, D. L. Sounas, A. Alu, A. Toscano, and F. Bilotti, "Phase-Induced Frequency Conversion and Doppler Effect with Time-Modulated Metasurfaces," *IEEE Trans. Antennas Propag.*, vol. 68, no. 3, pp. 1607–1617, Mar. 2020.

[16] L. Zhang *et al.*, "Space-time-coding digital metasurfaces," *Nat. Commun.*, vol. 9, no. 1, p. 4334, Dec. 2018.

[17] Z. Wu and A. Grbic, "Serrodyne Frequency Translation Using Time-Modulated Metasurfaces," *IEEE Trans. Antennas Propag.*, vol. 68, no. 3, pp. 1599–1606, Mar. 2020.

[18] M. Liu, D. A. Powell, Y. Zarate, and I. V. Shadrivov, "Huygens' Metadevices for Parametric Waves," *Phys. Rev. X*, vol. 8, no. 3, p. 031077, Sep. 2018.

[19] X. Guo, Y. Ding, Y. Duan, and X. Ni, "Nonreciprocal metasurface with space–time phase modulation," *Light Sci. Appl.*, vol. 8, no. 1, pp. 1–9, Dec. 2019.

[20] Y. Hadad, D. L. Sounas, and A. Alu, "Space-time gradient metasurfaces," *Phys. Rev. B*, vol. 92, no. 10, p. 100304, Sep. 2015.

[21] S. J. Orfanidis, *Electromagnetic Waves and Antennas*. https://www.ece.rutgers.edu/~orfanidi/ewa/.

[22] M. M. Salary, S. Jafar-Zanjani, and H. Mosallaei, "Electrically tunable harmonics in time-modulated metasurfaces for wavefront engineering," *New J. Phys.*, vol. 20, no. 12, p. 123023, Dec. 2018.

[23] D. L. Sounas, C. Caloz, and A. Alù, "Giant non-reciprocity at the subwavelength scale using angular momentum-biased metamaterials," *Nat. Commun.*, vol. 4, no. 1, p. 2407, Dec. 2013.

[24] "CST Studio Suite 3D EM simulation and analysis software." [Online]. Available: https://www.3ds.com/products-services/simulia/products/cst-studio-suite/?utm_source=cst.com&utm_medium=301&utm_campaign=cst. [Accessed: 05-Apr-2020].

[25] S. Taravati and A. A. Kishk, "Advanced Wave Engineering via Obliquely Illuminated Space-Time-Modulated Slab," *IEEE Trans. Antennas Propag.*, vol. 67, no. 1, pp. 270–281, Jan. 2019.